\documentclass[a4paper,12pt]{article}

\usepackage{amsthm,amsmath}
\usepackage{natbib}
\usepackage[colorlinks,citecolor=blue,urlcolor=blue]{hyperref}
\usepackage{amssymb}
\usepackage{graphicx}
\usepackage{amsfonts}
\usepackage{amsbsy}
\usepackage{mathrsfs}
\usepackage{lscape}
\usepackage{float}
\theoremstyle{plain}

\usepackage{multirow}
\usepackage{tablefootnote}
\usepackage{threeparttable}
\usepackage{pdflscape}
\usepackage{enumerate}
\usepackage{pdfpages}
\usepackage{authblk}
\usepackage{siunitx}

\usepackage{caption}
\usepackage{subcaption}

\providecommand{\noopsort}[1]{}
\numberwithin{equation}{section}
\theoremstyle{plain}

\usepackage[english]{babel}

\newcommand{\de}{\mathrm{d}}

\begin{document}
\graphicspath{{pics/}{pics/}}

\title{On adaptive kernel intensity estimation on linear networks}
\author[1]{Jonatan A. Gonz\'alez\footnote{(corresponding author) Email: jonathan.gonzalez@kaust.edu.sa}}
\author[2]{Paula Moraga}
\affil[1,2]{Computer, Electrical and Mathematical Science and Engineering Division\\
King Abdullah University of Science and Technology (KAUST)\\
Thuwal 23955-6900, Saudi Arabia}
\maketitle

\begin{abstract}
In the analysis of spatial point patterns on linear networks, a critical statistical objective is estimating the first-order intensity function, representing the expected number of points within specific subsets of the network. Typically, non-parametric approaches employing heating kernels are used for this estimation. However, a significant challenge arises in selecting appropriate bandwidths before conducting the estimation. We study an intensity estimation mechanism that overcomes this limitation using adaptive estimators, where bandwidths adapt to the data points in the pattern. While adaptive estimators have been explored in other contexts, their application in linear networks remains underexplored. We investigate the adaptive intensity estimator within the linear network context and extend a partitioning technique based on bandwidth quantiles to expedite the estimation process significantly. Through simulations, we demonstrate the efficacy of this technique, showing that the partition estimator closely approximates the direct estimator while drastically reducing computation time. As a practical application, we employ our method to estimate the intensity of traffic accidents in a neighbourhood in Medellin, Colombia, showcasing its real-world relevance and efficiency.
\end{abstract}


\section{Introduction}\label{sec:introduction} Spatial point processes on linear networks are a variation of the spatial point process framework that accounts for the geometry of a linear network in which events occur \citep{ang2012linearNetworks}. In this framework, events are characterised by their spatial location along the network. This spatial information can be used to understand the relationships between events and the network topology and to model how the linear network influences the occurrence of events.

For example, consider the analysis of traffic accidents on a highway. The occurrence of an accident is not only determined by factors such as driver behaviour and weather conditions but also by the spatial characteristics of the road itself, such as its curves, intersections, and lane configurations. Using a point process on linear spatial networks makes it possible to model the occurrence of accidents as a function of the location and investigate how specific road features affect the likelihood of an accident \citep{baddeley2015spatialR}.

Point processes on linear spatial networks have numerous applications in transportation engineering, epidemiology, and ecology, where events occur on linear features such as roads, rivers, and pipelines \citep{baddeley2021reviewNetworks}. Examples of line networks include maps of railways, rivers, electrical wires, nerve fibres, airline routes, irrigation canals, geological faults, or soil cracks. Data points can be associated with traffic accidents, street crimes, roadside trees, retail stores, roadside kiosks, insect nests, neuroanatomical features, or sample points along a stream. Even John Snow's pioneering work on cholera cases in London can be considered a point pattern on a linear network representing the district streets \citep{baddeley2021reviewNetworks}. However, the statistical analysis of network data presents significant challenges. The non-homogeneity of network data creates computational and geometrical complexities, leading to new methodological problems that may generate methodological errors.

Moreover, the spatial scales of network data can vary widely, further complicating the analysis. Such challenges pose a significant and far-reaching problem to the classical methodology of spatial statistics based on stationary processes, which is mainly inapplicable to network data. Nevertheless, analysing point patterns on linear networks has gained increased attention from GIS and spatial statistics communities in recent years. In spatial ecology and spatial statistics, geostatistical techniques have been focused on analysing spatial variables on a network of rivers or streams.

\begin{figure}[h!t]
    \centering
    \includegraphics[width=1\linewidth]{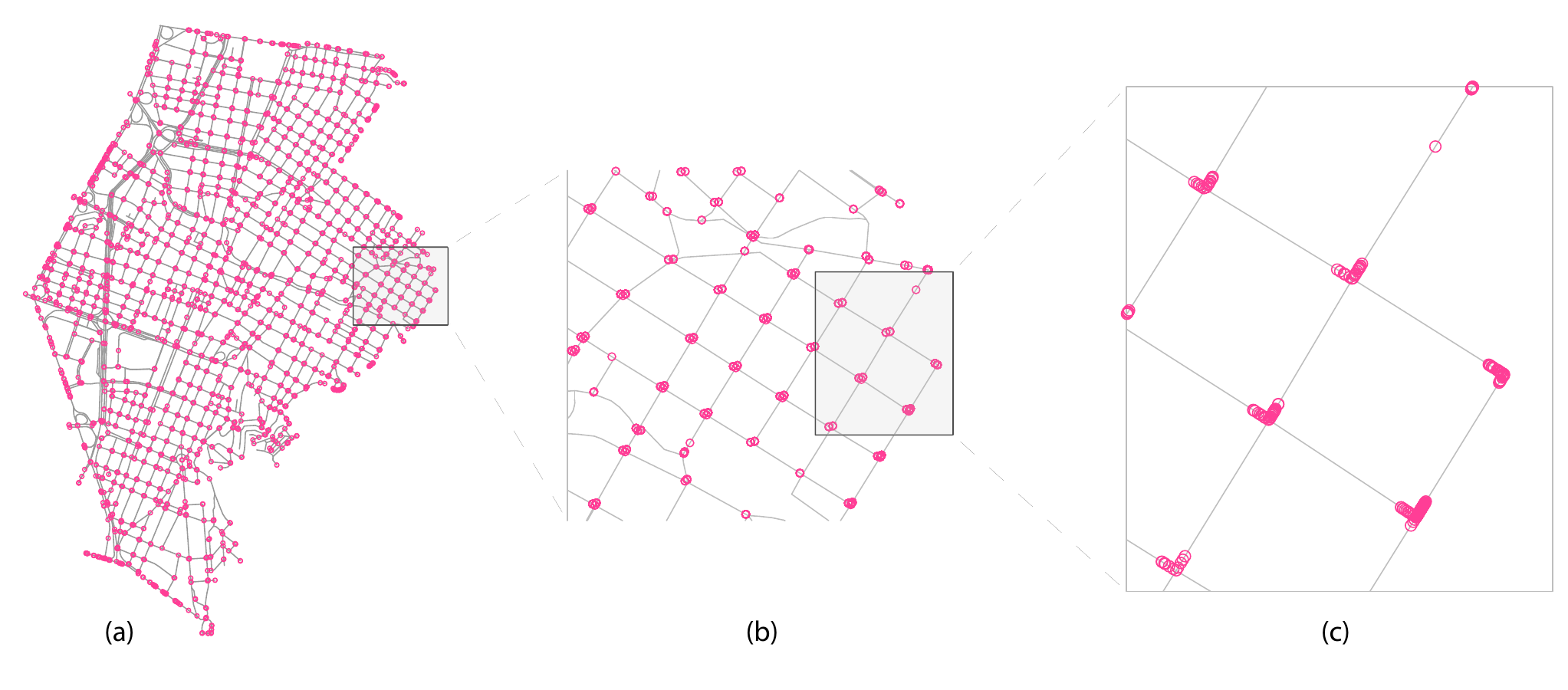}
    \caption{(a): Traffic accidents (pink dots) in the road network of Candelaria's neighbourhood of Medellin, Colombia (Comuna 10) $(X_{L_0})$ from 2014 to 2019, a total of 76610 data points (traffic accidents). (b): An enlargement from Candelaria's east part $(X_{L_1})$. (c): A second enlargement of Candelaria's zone $(X_{L_2})$.}
    \label{fig:candelarialpp}
\end{figure}

Candelaria is an emblematic neighbourhood within Medell\'in, Colombia. The traffic patterns within Candelaria address mobility demands, often manifesting as a rhythmic interweaving of vehicular and pedestrian activities. The data representing Candelaria's road accidents (a point pattern denoted by $X_{L_0}$ hereinafter, where $L_0$ represent Candelaria's road network), as depicted in Figure \ref{fig:candelarialpp}(a), demonstrate significant spatial disparities in the distribution of data points. In such situations, the conventional approach of employing fixed-bandwidth kernel estimation often proves inadequate. Let us define $X_{L_1}$ and $X_{L_2}$ as two subsets of $X_{L_0}$, such that $X_{L_2}\subset X_{L_1}\subset X_{L_0}$, where $L1$ and $L_2$ are nested subsets of Candelaria's road network $L_0$. Precisely, dense clusters of accidents occurring at certain intersections (see Figure \ref{fig:candelarialpp}(b), (c)) within the urban road network may be excessively smoothed out. In contrast, sparser accident occurrences in other areas of the road system may be insufficiently smoothed. To address this challenge, implementing adaptive kernel estimation, which allows for variable bandwidth, can yield considerably superior results \citep{abramson1982bandwidth, Davies2018kernel}.

Furthermore, additional complexity arises from the fact that the road network itself exhibits extensive spatial variation. This characteristic adds further difficulty to visually assessing the accident density per unit length of road. Moreover, this variation poses challenges from a computational standpoint due to the requirement of a much finer spatial resolution in certain intersections compared to straight-road segments. Consequently, accurately capturing and analyzing accident patterns necessitates a nuanced and adaptable approach that accounts for both the variability in accident concentrations and the spatial intricacies of the road network.

This paper's objectives are twofold: First, to expand the adaptive methodology for intensity estimation via kernel in the context of point patterns on networks, specifically focusing on the heat kernel. Secondly, we extend a partition algorithm to estimate within reasonable timeframes. In addition, the method has been implemented in the R package \texttt{kernstadapt} \citep{kernstadaptpackage}.

Our paper is structured as follows: Section \ref{sec:fundamentals} lays down the foundational concepts of point patterns on linear networks. Section \ref{sec:intensityfunction} delves into introducing and estimating the intensity function, employing various methodologies encompassing our central approach - the heat kernel. Moving forward, Section \ref{sec:adaptiveestimators} introduces adaptive estimators for intensity estimation by kernels. Section \ref{sec:partitioningalgorithm} introduces the partition algorithm designed to expedite the computational process of adaptive estimators, accompanied by simulation demonstrations of our method's efficacy. In Section \ref{sec:application}, we apply our developed technique to estimate the intensity of traffic accidents in Candelaria. Finally, in Section \ref{sec:discussion}, we discuss our findings while presenting novel avenues for future research exploration.

\section{Fundamentals: Point processes on linear networks}\label{sec:fundamentals}
A {\it linear network} can be defined as the union of a finite number of line segments, denoted by $L$. Let $l_i$ be a line segment; then, it takes the parametrised form
$$
l_i=[\mathbf{u}_i,\mathbf{v}_i]=\left\{w:w=t \mathbf{u}_i + (1-t)\mathbf{v}_i, 0\leq t \leq 1 \right\},
$$
where $\mathbf{u}_i$ and $\mathbf{v}_i$ are the endpoints. It is assumed that the intersection between two different segments is empty or an endpoint of both segments. The total length of a subset $B$ of the linear network is denoted by $|B|$. 

A {\it point pattern} on a linear network is a finite unordered set $X=\left\{\mathbf{x}_i \right\}_{i=1}^n$, where each point represents a location on the linear network. An observed point pattern is a realisation of a random point process $\mathbf{X}$ on the linear network. We assume that the number of points is finite, has finite mean and variance, and there are no multiple coincident points \citep{daley2003,baddeley2021reviewNetworks}.

We can measure distances in a linear network through the {\it shortest path}. A {\it path} between two points $\mathbf{u}$ and $\mathbf{v}$ in $L$ is a points sequence $\left\{\mathbf{y}_i \right\}_{j=0}^m$, such that $\mathbf{y}_0=\mathbf{u}$ and $\mathbf{y}_m=\mathbf{v}$, and every segment line $[\mathbf{y}_j,\mathbf{y}_{j+1}]\subset L$, for $j=0,\ldots,m-1$. If $||\cdot ||$ denotes the Euclidean distance, the length of the path $\left\{\mathbf{y}_j \right\}_{j=0}^m$ is given by 
$$
\sum_{j=0}^{m-1}||\mathbf{y}_{j+1} - \mathbf{y}_j||.
$$
The {\it shortest path distance} $d_{\ell}(\mathbf{u},\mathbf{v})$ between $\mathbf{u}$ and $\mathbf{v}$ in a linear network $L$ is defined as the minimum of the lengths of all paths from $\mathbf{u}$ to $\mathbf{v}$. The distance is infinite if no paths are from $\mathbf{u}$ to $\mathbf{v}$. 

In the shortest-path metric, a disc with radius $r>0$ and centre $\mathbf{u}$ in the network $L$  is the set of all points lying no more than a distance 
$r$ from the location $\mathbf{u}$, in the shortest path distance; formally, 
$$
b_L(\mathbf{u},r)=\left\{\mathbf{v}\in L: d_L(\mathbf{u}, \mathbf{v})\leq r \right\}.
$$

\section{Intensity function}\label{sec:intensityfunction}
Given the data points of a point pattern $X$ in a network $L$ conceived as a realisation of the point process $\mathbf{X}$, we want to estimate the intensity function $\lambda(\mathbf{u})$.
Let $N(B)$ denote the number of points of $\mathbf{X}\cap B$ in a subset $B$ of the linear network. Then the point process has intensity function $\lambda(\mathbf{u}),\mathbf{u}\in L$ if for all closed subsets
\begin{equation}
    \mathbb{E}\left[ N(B) \right] = \int_B \lambda(\mathbf{u}) \de_{\ell}\mathbf{u},  
\end{equation}
where the integration is done with respect to arc length on the linear network. The intensity function may be interpreted as the expected number of points per unit length.

Campbell’s formula \citep[][section 13.1, p. 269]{daley2007} applies for linear networks and it is given by
\begin{equation*}
    \mathbb{E}\left[ \sum_{\mathbf{u}_i \in \mathbf{X}}h(\mathbf{u}_i) \right] = \int_L h(\mathbf{u}) \lambda(\mathbf{u}) \de_{\ell}\mathbf{u},
\end{equation*}
where $h$ is a real, measurable function. 


\subsection{Estimators}
A straightforward first step in analysing spatial point patterns is to estimate the intensity function using kernels \citep{baddeley2015spatialR}. This option suits linear networks but has many more considerations \citep{mcswiggan2017kernelnetworks}.
We start by considering a kernel estimator of the form
\begin{equation}\label{eq:kerneletimator}
    \hat{\lambda}(\mathbf{u}) = \sum_{i=1}^n K (\mathbf{u}|\mathbf{u}_i), \quad \mathbf{u}\in L,
\end{equation}
where $K$ is a smoothing kernel on the real line that must satisfy some properties. It should be non-negative, and it should have a total mass of 1, i.e., $\int_L K(\mathbf{u}|v)\de_{\ell} \mathbf{v}=1$ for all $\mathbf{u}\in L$. Then, the mass of the intensity estimator must be $n$, the total number of points. 
A classical estimator of the type given in Eq.\eqref{eq:kerneletimator}, where $K (\mathbf{u}|\mathbf{u}_i)=\kappa(d_{\ell}(\mathbf{u},\mathbf{u}_i))$, and where $\kappa$ is a smoothing kernel on the real line will not preserve the mass \citep{mcswiggan2017kernelnetworks}. Therefore, other options must be considered to reduce the bias.

\subsubsection{Edge-corrected estimators}
One way to correct the bias of the ``natural estimator'' is considering {\it edge correction factors} intended for preserving the mass \citep{moradi2017Kernel, baddeley2021reviewNetworks}. For a point pattern $X$ on a linear network, consider an edge correction function given by 
\begin{equation*}
    c_L(\mathbf{u}):= \int_{L} \kappa(d_{\ell}(\mathbf{u},\mathbf{v})) \de_{\ell} \mathbf{v}.
\end{equation*}
Then, an intensity estimator $\hat{\lambda}^{\text{U}}(\mathbf{u})$ based on the uniform edge correction \citep{diggle1985kernel} and one based on \citeauthor{Jones1993}'s (\citeyear{Jones1993}) edge correction may be defined as
\begin{equation*}
    \hat{\lambda}^{\text{U}}(\mathbf{u}) =\frac{1}{c_L(\mathbf{u})} \sum_{i=1}^n \kappa(d_{\ell}(\mathbf{u},\mathbf{u}_i)),
    \quad 
    \text{and}
    \quad
    \hat{\lambda}^{\text{JD}}(\mathbf{u}) = \sum_{i=1}^n \frac{\kappa(d_{\ell}(\mathbf{u},\mathbf{u}_i))}{c_L(\mathbf{u}_i)}.
\end{equation*}
The computation of the edge-correction factors $\hat{\lambda}^{\text{U}}(\mathbf{u})$ and $\hat{\lambda}^{\text{JD}}(\mathbf{u})$ can be computationally burdensome \citep{baddeley2021reviewNetworks}.  Given this significant computational cost, exploring alternative methods when seeking a rapid estimation is advisable. 

\subsubsection{Equal-split kernel estimators}
\cite{okabe2012spatial}, Chap. 9, and reference therein, summarise kernel density estimators on a general network. Their work investigated computational algorithms that redistribute the mass of a kernel $\kappa$ from the real line onto the network. They identified desirable properties for a kernel estimator and found that the so-called {\it equal-split discontinuous} and {\it equal-split continuous} rules satisfied many of them.

The rule known as {\it continuous} possesses remarkable features such as symmetry, mass preservation, and unbiasedness under the assumption of true uniform intensity. On the other hand, the edge-corrected estimator $\hat{\lambda}^{\text{U}}(\mathbf{u})$ is unbiased but does not preserve mass, whereas $\hat{\lambda}^{\text{JD}}(\mathbf{u})$ preserves mass but is not unbiased, and neither of them is symmetric. Regrettably, implementing the ``continuous'' rule algorithm is exceedingly slow \citep{mcswiggan2017kernelnetworks}. While the {\it discontinuous} rule is faster, it is associated with less favourable characteristics as a non-continuous estimate of the intensity \citep[][Sec. 9.3.2]{okabe2012spatial}. Both methods utilise a kernel on the real line with bounded support on a network, which precludes using the Gaussian kernel. The computational burden increases exponentially with the bandwidth, making automatic bandwidth selection computationally impractical \citep{baddeley2021reviewNetworks}.

The algorithm {\it equal-split discontinuous} generates a replica of the kernel $\kappa$ for each point of $X$. For locations $\mathbf{u}$ that share a line segment with the origin $\mathbf{u}_i$, the kernel estimate takes the value $\kappa(d_{\ell}(\mathbf{u},\mathbf{u}_i))$. Whenever the network branches out, the residual tail mass of the kernel is distributed uniformly among the new line segments, ensuring that the total mass is conserved.

When a network lacks loops, let $\mathbf{u}$ be a location of the network $L$, $\mathbf{u}_i$ a data point, and $\{m_i\}_{j=1}^p$ be the degrees (the number of edges incident to a vertex) of each vertex along the shortest path from $\mathbf{u}_i$ to $\mathbf{u}$ without $\mathbf{u}$ and $\mathbf{u}_i$. Then, the equal-split discontinuous kernel is given by
\begin{equation*}
    K^{D}(\mathbf{u}|\mathbf{u}_i)=\frac{\kappa(d_{\ell}(\mathbf{u},\mathbf{u}_i))}{\prod_{j=1}^p (m_j - 1)}.
\end{equation*}
When there are possible loops, let $\pi^*=(\mathbf{u}_i,\mathbf{v}_1, \ldots, \mathbf{v}_{P-1}, \mathbf{u})$ denote the paths, from $\mathbf{u}_i$ to $\mathbf{u}$, of length less than or equal to $h$. These paths are {\it non-reflecting}, i.e., $\mathbf{e}_i\neq \mathbf{e}_{i+1}$, with $\mathbf{e}_i$ as the edge containing $\mathbf{v}_{i-1}$ and $\mathbf{v}_i$. Let $\ell(\pi)$ denote the length of the path, and 
$$
a^D(\pi):=\frac{1}{\prod_{j=1}^{P-1}(m_j-1)},
$$ 
where $m_j$ is the degree of $\mathbf{v}_j$. Thus, the equal-split discontinuous kernel is given by 
\begin{equation*}
    K^{D}(\mathbf{u}|\mathbf{u}_i)=\sum_{\pi^*} \kappa(\ell(\pi)) a^D (\pi).
\end{equation*}
The {\it equal-split continuous} kernel estimator corresponds to modifying the previous algorithm to generate a continuous function on the network \citep{okabe2012spatial,mcswiggan2017kernelnetworks,baddeley2021reviewNetworks}. This modified approach extends to paths of length less than $h$, including those that reflect at vertexes. When a path arrives at a vertex of degree $m$, it encounters $m-1$ outgoing branches and one incoming branch. Assuming that a weight $2/m$ is assigned to each outgoing branch, and a weight $(2/m - 1)$ is given to the incoming branch. The kernel achieves continuity, and due to the assumption of monotonicity of the kernel $\kappa$, the resulting function maintains non-negative values.

The equal-split continuous kernel estimator can then be written by 
\begin{equation*}
    K^{C}(\mathbf{u}|\mathbf{u}_i)=\sum_{\pi} \kappa(\ell(\pi)) a^C (\pi),
\end{equation*}
where
$$
a^C(\pi):=\prod_{j=1}^{P-1}\left(\frac{2}{\text{deg}(\mathbf{v}_j)} - \delta_j \right),
$$ 
and where $\delta_1=\delta_P=0$, and $\delta_j = \mathbf{1}\left\{ \mathbf{e}_j = \mathbf{e}_{j-1}\right\}, j=2,\ldots,P-1$.

\subsubsection{Heat kernel}
\cite{mcswiggan2017kernelnetworks} introduced a statistically rigorous kernel estimator for a linear network. They achieved this by leveraging the relationship between kernel smoothing and diffusion, as established by previous works such as \cite{chaudhuri2000scalespace} and \cite{botev2010kerneldifussion}. When applied to a network, the heat kernel is equivalent to the Gaussian kernel, representing the function that describes the diffusion or spread of heat across the network, similar to its role in classical physics.

We first consider the case of the real line. Brownian motion on an infinite straight line can be described as a stochastic process, denoted as $\{X(t)\}_{t\geq 0}$. In this process, the increments $X(t_j)-X(t_{j-1}),j=2\ldots,k$ between any two temporal points, $0\leq t_1<t_2<\ldots < t_k$, are independent Gaussian random variables with a mean zero and variances $t_j - t_{j-1}$. If we consider a Brownian motion $\{X(t)\}_{t\geq 0}$ that starts at a specific position $x_0$, the probability density of $X(t)$ later, say at time $t$, can be characterised by a Gaussian distribution with mean $x_0$ and a variance $\sigma^2 = t$. In the case of a Brownian motion that starts at a random position $x_0$ with probability density function $p(x)$, the probability density of $X(t)$ is given by
\begin{equation*}
    f_t(x)=\int_{-\infty}^{\infty} p(u) \kappa_t(x-u)\de u,
\end{equation*}
where $\kappa_t(\cdot)$ represents the Gaussian probability density function with mean zero and variance $t$. The function $f_t(x)$ also represents the solution to the traditional {\it heat equation} given by
\begin{equation*}
    \frac{\partial f}{\partial t}=\beta \frac{\partial ^2 f}{\partial x^2},
\end{equation*}
with {\it thermal diffusivity constant} $\beta$ and initial condition $f_0(x)=p(x)$. The solution to the heat equation can be represented by a kernel operator with kernel $\kappa_t$. In the case of the {\it heat kernel} on the real line, it is given by $\kappa_t$. Consequently, the standard Gaussian kernel estimator for a set of data points $\{x_i\}$ on the real line can be obtained by summing the values of the heat kernel $\kappa_t(x-x_i)$.

On the other hand, Brownian motion on a linear network is a specific type of diffusion occurring on a graph. It is a continuous-time Markov process, denoted as $\{X(t)\}_{t\geq 0}$, equivalent to one-dimensional Brownian motion on each network segment \citep[see, e.g., \ ][and references therein]{mcswiggan2017kernelnetworks}. Whenever the process reaches a vertex on the network with degree $m$, it has an equal likelihood of continuing along any of the $m$ edges connected to that vertex (including the edge it arrived from). Notably, if it reaches a terminal endpoint, it is instantaneously reflected \citep{mcswiggan2017kernelnetworks}.

The probability density function of Brownian motion on a linear network at time $t$, denoted as $f_t(\mathbf{u}),\mathbf{u}\in L$, obeys the classical heat equation on the network. This means that the analogous form of the classical heat equation holds at any position $\mathbf{u}$ that is not a vertex. It is important to note that the second spatial derivative is well-defined, irrespective of the chosen local coordinates on the line segment. Define the first spatial derivative of $f$ at $\mathbf{v}$ in the direction towards $\mathbf{v}'$ as
$$
\left.\frac{\partial f}{\partial \mathbf{u}[\mathbf{v},\mathbf{v}']}\right|_{\mathbf{v}}:= \lim_{h \downarrow 0}\frac{f(\mathbf{v}+h(\mathbf{v}-\mathbf{v}'))}{h||\mathbf{v}-\mathbf{v}'||}.
$$
At any vertex $\mathbf{v}$, the density $f_t$ is continuous and holds a property akin to the conservation of heat flow that can be expressed as follows,
\begin{equation}\label{eq:conservationcondition}
    \sum_{\mathbf{v}'\sim \mathbf{v}}\left.\frac{\partial f}{\partial \mathbf{u}[\mathbf{v},\mathbf{v}']}\right|_{\mathbf{v}}=0,
\end{equation}
where the sum is indexed over all edges $[\mathbf{v},\mathbf{v}']$ incident at vertex $\mathbf{v}$. Assume that we have a given initial condition $f_0(\mathbf{u}) = p(\mathbf{u})$ and the heat kernel $\kappa_t(\mathbf{u}|\mathbf{s})$ on the network. A solution to the heat equation and the conservation condition for heat flow on the linear network given in Eq. \eqref{eq:conservationcondition} can be represented as a kernel operator as follows,
$$
f_t(\mathbf{u})=\int_L p(\mathbf{s}) \kappa_t(\mathbf{u}|\mathbf{s})\de_{\ell} \mathbf{s}.
$$
In intuitive terms, $\kappa_t(\mathbf{u}|\mathbf{s})\de_{\ell} \mathbf{s}$ represents the probability that a Brownian motion on the network, initiated at position $\mathbf{s}$ at time $0$, will land within the tiny interval of length $\de_{\ell} \mathbf{s}$ around the point $\mathbf{u}$ at time $t$. Alternatively, we can interpret $\kappa_t(\mathbf{u}|\mathbf{s})$ as the transfer function that connects the temperature at location $\mathbf{s}$ at time $0$ to the temperature at location $\mathbf{u}$ at time $t$. It is worth noting that condition \eqref{eq:conservationcondition} implies that the first spatial derivative of $f$ must be zero at all terminal endpoints. In a physical sense, this assumption indicates that the network is thermally isolated, preventing heat from escaping, including from the terminal endpoints \citep{mcswiggan2017kernelnetworks}.

Estimating the intensity through a {\it diffusion intensity estimator}, $\hat{\lambda}^H(\mathbf{u})$, can be achieved mathematically by employing a sum of heat kernels, i.e., 
\begin{equation*}
    \hat{\lambda}^H(\mathbf{u})=\sum_{i=1}^n \kappa_t(\mathbf{u}|\mathbf{u}_i), \quad \mathbf{u}\in L,
\end{equation*}
where the time parameter is the squared bandwidth, i.e., $t=\sigma^2$. Note that this mathematical expression should not be directly employed in computations. Instead, the diffusion estimator can be obtained by numerically solving the time-dependent heat equation up to the desired time. Numerical solutions of the heat equation are significantly faster \citep{mcswiggan2017kernelnetworks,baddeley2021reviewNetworks}, often by several orders of magnitude, compared to path-enumeration algorithms. The computational time increases quadratically with the bandwidth.

\section{Adaptive estimators}\label{sec:adaptiveestimators}
Adaptive estimators have been superficially studied in the case of linear networks \citep{baddeley2021reviewNetworks}. These have only been proposed in the case of corrected kernel sums \citep{rakshit2019fastlargenetworks}; i.e., sums of kernel functions.

In this work, we want to go further and provide techniques for adaptive estimation that cover all cases, with particular emphasis on the heat kernel estimator. The general rationale is the following: a kernel estimator of the form given in Eq. \eqref{eq:kerneletimator} usually is equipped with a bandwidth $\epsilon$. This bandwidth may be a positive constant representing the standard deviation of the kernel involved or a variance-covariance matrix in the anisotropic case \citep{baddeley2015spatialR}. We consider the isotropic case, but instead of constant, the bandwidth will be a spatially varying function $\epsilon(\mathbf{u}), \mathbf{u}\in L$. Therefore, the intensity estimator takes the general form
\begin{equation*}
    \hat{\lambda}_{\epsilon}(\mathbf{u}) = \sum_{i=1}^n K_{\epsilon(\mathbf{u}_i)}(\mathbf{u}|\mathbf{u}_i), \quad \mathbf{u}\in L,
\end{equation*}
where $\epsilon(\mathbf{u})$ is a bandwidth function defined as
\begin{equation}\label{eq:abramsombandwidth}
\epsilon(\mathbf{u})=\frac{\epsilon^{\star}}{\gamma} \sqrt{\frac{n}{\tilde{\lambda}(\mathbf{u})}},  \quad \mathbf{u}\in L,
\end{equation}
where $\epsilon^{\star}$ is a spatial smoothing multiplier known as \textit{global bandwidth}, $\tilde{\lambda}(\mathbf{u})$ is a fixed-bandwidth pilot estimate of the intensity function (estimated using the global bandwidth), and $\gamma$ is the geometric mean terms for the marginal intensities evaluated in the points of the point pattern, i.e.,
$$
\gamma := \exp \left \{ \frac{1}{n} \sum_{i=1}^n \log\{\lambda{(\mathbf{u}_i)^{-2}}\}\right\}
$$
This approach was proposed originally by \cite{abramson1982bandwidth}; the inclusion of the geometric mean frees the bandwidth from the data scale \citep{silverman1986density,davieshazelton2010adaptiverisk}. 

\begin{figure}[h!tb]
\centering
\begin{subfigure}{.48\linewidth}
\centering
\includegraphics[width=0.8\linewidth]{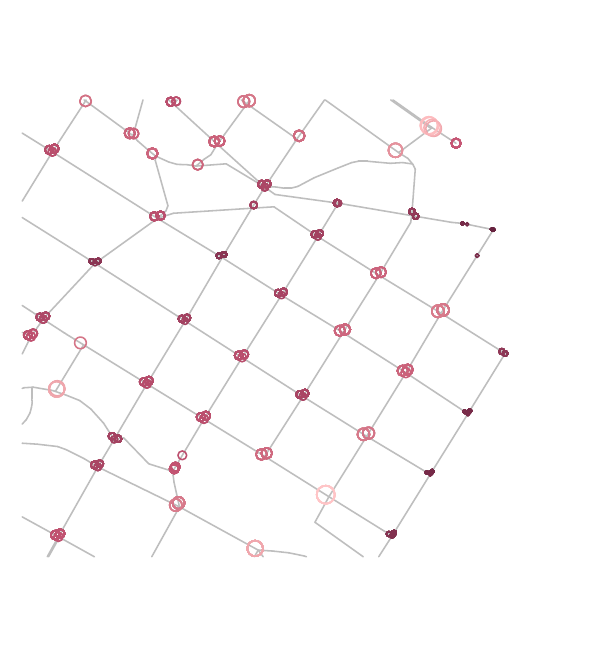}
\caption{}
\end{subfigure}
\begin{subfigure}{.4\linewidth}
\centering
\includegraphics[width=0.8\linewidth]{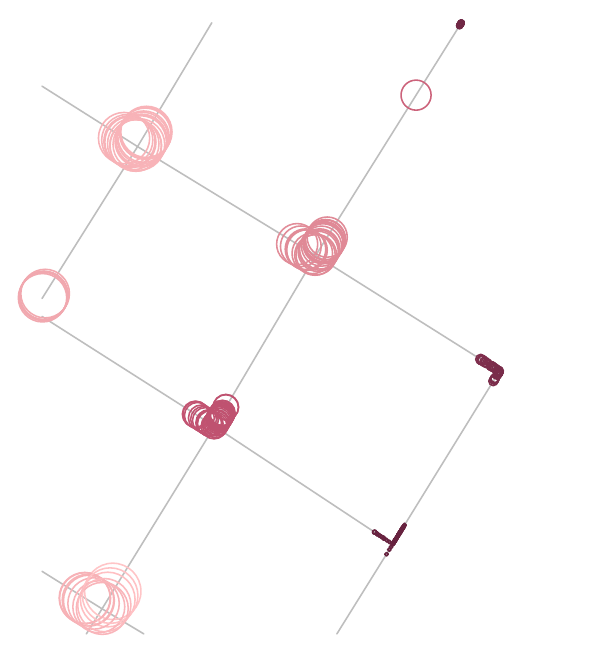}
\caption{}
\end{subfigure}
\caption{Candelaria's traffic accidents point patterns (enlargements $X_{L_1}$ in (a) and  $X_{L_2}$ in (b)) with adaptive kernel bandwidth from Abramson's rule. Smaller and darker dots represent shorter bandwidths, i.e., crowded regions in $L_1$ and $L_2$.}
\label{fig:variablebandwidth}
\end{figure}

\section{Partition algorithm}\label{sec:partitioningalgorithm}
The partition algorithm is a method used in point processes to determine adaptive bandwidths \citep[see, e.g.,\ ][]{Davies2018kernel}. The partition algorithm aims to find suitable bandwidth values for kernel smoothing. As the adaptive bandwidth approach recognizes that the density of points can vary across the spatial domain, it seeks to assign smaller bandwidths in more crowded regions and larger bandwidths in less overcrowded areas. 

By adapting the bandwidth values to the local density of points, the partition algorithm allows for a more accurate and flexible intensity estimation. It captures the spatial heterogeneity of the point pattern according to a prespecified number of bins.

The partition algorithm works as follows: Initially, a candidate bandwidth $\hat h_i:= \epsilon(\mathbf{u}_i)$ is computed for every point $\mathbf{u}_i$ of the point pattern. The candidate is obtained through the \cite{abramson1982bandwidth} formula by Eq. \eqref{eq:abramsombandwidth}.

Then we consider the empirical $\zeta$th quantiles, say $\hat h^{(\zeta)}$ of the $n$ bandwidths. Then we consider a {\it quantile step} $0<\delta \leq 1$, such that $D= 1/\delta$ is an integer; in practice, $\delta \ll 1$. Then we define a set of bins employing the sequence of values $\hat h^{(0)},\hat h^{(\delta)}, \ldots, \hat h^{(1)}$ such that every observation $\mathbf{u}_i$ belongs to one of the bins
\begin{equation*}
    \left[h^{(0)}, h^{(\delta)}\right], \left(h^{(\delta)}, h^{(2 \delta)}\right], \ldots, \left( h^{((D-1)\delta)}, h^{(1)}\right].
\end{equation*}
If $X_d$ is the point pattern consisting of the points that belong to the $d$th bin, then 
$$
X=\bigcup_{d=1}^D X_d.
$$
To approximate the adaptive smoother, we will substitute each desired bandwidth $\hat h_i$ with the bin's midpoint. The approximation results from summing the $D$ fixed-bandwidth estimates applied to the respective subsets of the original point pattern, i.e.,
\begin{equation}\label{eq:partitioninglambda}
	\hat{\lambda}_{\epsilon}\left(\mathbf{u}\right) \approx \sum_{d=1}^{D} \hat{\lambda}^{*}_{\bar{\epsilon}_d}\left(\mathbf{u}| X_{d} \right), \quad \mathbf{u}\in L,
\end{equation}
where $\bar{\epsilon}_i$ is the midpoint of the $i$th bin, and $\hat{\lambda}^{*}_{\bar{\epsilon}_d}\left(\mathbf{u}| X_d \right)$ corresponds to a fixed-bandwidth intensity estimate of the sub-pattern $X_d$.

\subsection{Simulation study}
To explore the performance of the partition algorithm in networks by using the heating kernel, we launch a series of simulations designed to assess and contrast the outcomes of explicit calculations for adaptive kernel intensities with various partitioned versions. We aim to evaluate each approach's efficiency and accuracy by varying the parameters and configurations. Through these simulations, we seek to gain valuable insights into the trade-offs involved in adaptively estimating kernel intensities and the impact of employing partitioned estimators. Additionally, we measure the computation time for each method to gauge their practical feasibility and potential scalability.

While it can be theoretically argued that any kernel-type estimator possesses an adaptive version by defining a variable bandwidth, the practical implementation of such adaptivity is far from straightforward. The computational procedures involved can often become exceedingly complex. For instance, in the case of the equal-split kernel estimator, even with optimised code, computing it for a simple network can be time-consuming. When considering adaptive computation, where each point requires a total computation of the estimator (or several sets of points when using the partition algorithm), the complexity becomes prohibitive, rendering the computation impossible or incredibly slow. Given these challenges, in this study, we deliberately decided to exclude the ``Equally-split-kernel'' estimator from practical use. Instead, we focused on the heating kernel as the most viable and effective option for adaptive estimation thus far. By doing so, we acknowledge the limitations in adapting some kernel-type estimators while highlighting the heating kernel's sufficiency for practical adaptive intensity estimation.

In our simulations, the domain $L_2$ (a subset of Candelaria's road network as defined in Section \ref{sec:introduction}) has been deliberately chosen as the basis for analysis, corresponding to an enlarged representation of Candelaria's neighbourhood. In contrast to planar point pattern methodology, linear networks lack a classical reference network akin to the unit square. Consequently, we must adopt an {\it ad hoc} approach in selecting a suitable linear network for effectively demonstrating our methods. By opting for $L_2$, we ensure that the simulations are conducted within a context that closely mimics the characteristics and complexities of Candelaria's neighbourhood. In addition, we adjusted our simulations to have realisations of roughly 520 points.

\subsection{Log-Gaussian restricted to linear networks}
This is the restriction to the network of a log-Gaussian random field with zero mean and an exponential correlation function with some variances and scales. In this case, the point patterns are observations of point processes with stochastic intensity $\Lambda(\mathbf{u})=\exp{(Z(\mathbf{u}))}$, where $Z(\mathbf{u})$ is a Gaussian random field observation \citep[see, e.g.,\ ][]{moller1998loggaussian}. 

\begin{figure}[h!tb]
\centering
\begin{subfigure}{.45\linewidth}
\centering
\includegraphics[width=0.99\linewidth]{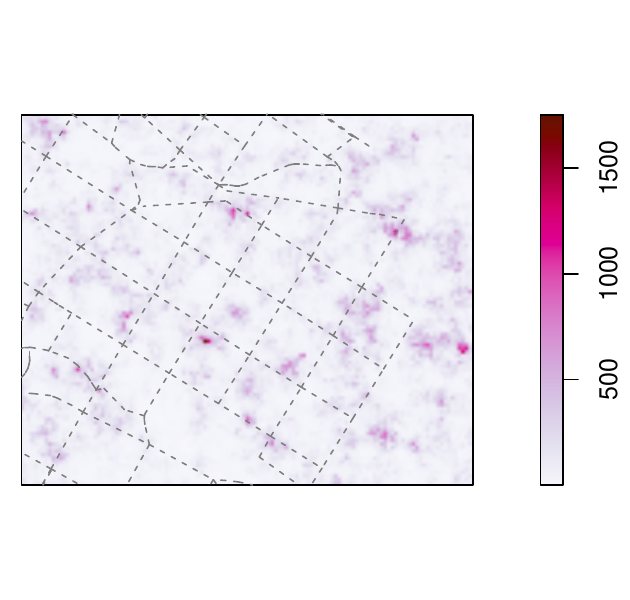}
\caption{}
\end{subfigure}
\begin{subfigure}{.45\linewidth}
\centering
\includegraphics[width=0.99\linewidth]{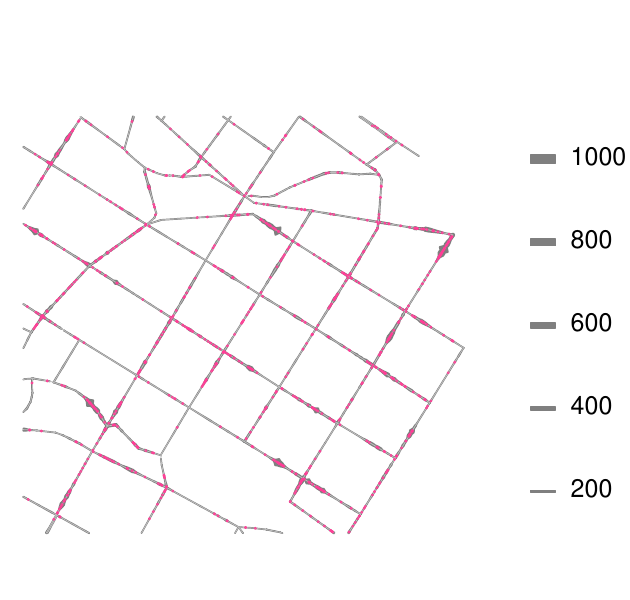}
\caption{}
\end{subfigure}
\caption{Simulated realisation of a point pattern in a linear network from log-Gaussian random field observation. (a) The random field observation is based on an initial two-dimensional surface defined within the unit square. (b) The point locations (pink) and the intensity visualisation (grey lines with variable width) in the linear network.}
\label{fig:loggaussian}
\end{figure}

We set a stationary exponential covariance function of the form
$$
\gamma(\mathbf{u}) = \sigma^2 \exp \left\{-\frac{||\mathbf{u}||}{\phi}\right\}, \quad \sigma^2,\phi \in \mathbb{R}_+,
$$
where the $\sigma^2$ and $\phi$ stand for the variance and scale, respectively. We select the following values for the variances and scales: $\sigma^2\in \{ 0.9, 2\}$ and $\phi \in \{0.03, 0.09\}$. So we consider 4 cases for this scenario: {\it logGaussian 1}: $\sigma^2=0.9, \phi = 0.03$, {\it logGaussian 2}: $\sigma^2=0.9, \phi = 0.09$, {\it logGaussian 3}: $\sigma^2=2, \phi = 0.03$ and {\it logGaussian 4}: $\sigma^2=2, \phi = 0.09$. Figure \ref{fig:loggaussian} shows an observation of this type of point pattern. The random field is generated on the unit square.

\subsection{Gaussian mixture}
\begin{figure}[h!tb]
\centering
\begin{subfigure}{.45\linewidth}
\centering
\includegraphics[width=0.99\linewidth]{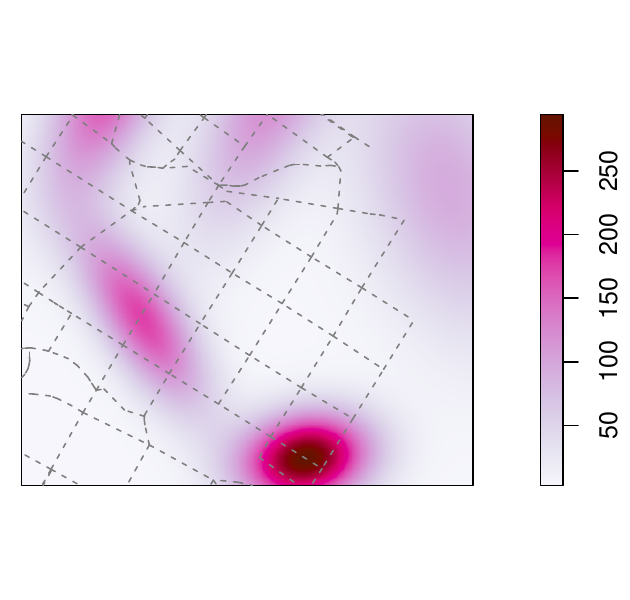}
\caption{}
\end{subfigure}
\begin{subfigure}{.45\linewidth}
\centering
\includegraphics[width=0.99\linewidth]{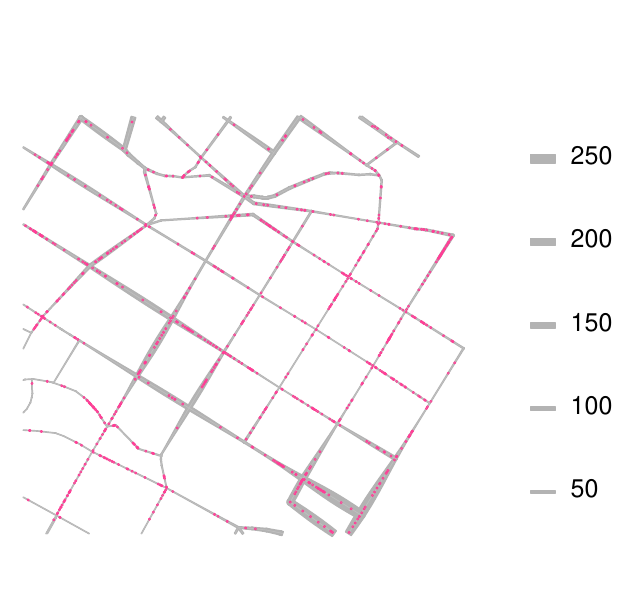}
\caption{}
\end{subfigure}
\caption{Simulated realisation of a point pattern in a linear network in Gaussian mixture fashion. (a) The Gaussian mixture realisation is based on an initial two-dimensional surface defined within the unit square. (b) The point locations (pink) and the intensity visualisation (grey lines with variable width) in the linear network.}
\label{fig:gaussianmixture}
\end{figure}

Given a set of $g$ spatial anisotropic Gaussian densities, each with mean vector $\mu_i, i\in g$ and covariance matrix $\Sigma_i$; we consider a mixture of them. A mixture of distributions is a probabilistic model that combines each with its own mean and covariance matrix to represent complex data with varying patterns and clusters. If $F_i(\mu_i, \Sigma_i)$ is the cumulative distribution function for every component, then the cumulative distribution function of the mixture is given by 
$$
F:=\sum_{i=1}^g w_i F_i(\mu_i, \Sigma_i),
$$
where $\sum_i w_i = 1$. In this case, we set uniform weights all equal to $1/5$. The mean vectors are randomly assigned so that all their components are observations of the uniform distribution inside the unit square. Then, the surface is evaluated only at locations on the linear network. In this case, we opt for five different distributions $(g=5)$ and manually assign the covariance matrices. 
\begin{equation*}
\Sigma_1 := 
\begin{bmatrix}
0.01   & -0.01 \\
-0.01 & 0.02 \\
\end{bmatrix}, \quad
\Sigma_2 := 
\begin{bmatrix}
0.016 & 0.02 \\
0.02 & 0.05 \\
\end{bmatrix}, \quad
\Sigma_3 := 
\begin{bmatrix}
0.01 & 0.01 \\
0.01 & 0.03 \\
\end{bmatrix},
\end{equation*}

\begin{equation*}
\Sigma_4 := 
\begin{bmatrix}
0.02   & -0.01 \\
-0.01 & 0.05 \\
\end{bmatrix}, \quad
\Sigma_5 := 
\begin{bmatrix}
0.01 & 0.001 \\
0.001 & 0.005 \\
\end{bmatrix}.
\end{equation*}
In a mixture of Gaussian distributions, the covariance matrices represent the spread and orientation of data points within each Gaussian component. They describe the relationships between different variables in the data and influence the shape of the distribution's ellipsoidal clusters. Figure \ref{fig:gaussianmixture} showcases a simulated realisation of a point pattern in a linear network, adopting the Gaussian mixture setting. The Gaussian mixture is generated based on an initial two-dimensional surface delineated within the confines of the unit square. The point pattern unveils the precise locations of the generated points within the linear network.

\subsection{Integrated squared error}
For a given point pattern on a linear network, the intensity estimation is carried out through a two-step process, starting with direct adaptive kernel estimation. Subsequently, partitioned estimates are executed, encompassing some possible values for the number of groups; indeed, we set $\delta \in \{0.1, 0.05, 0.025, 0.01\}$. A target resolution of $128\times 128$ pixels is adhered to throughout this analytical scenario, ensuring a detailed and granular examination of the pattern's intensity distribution. The accuracy of this approximation is methodically evaluated using the {\it integrated squared error} (ISE) as the key metric given by
$$
\text{ISE}[\hat{\lambda}]:= \int_L \left(\hat{\lambda}(\mathbf{u}) - \lambda(\mathbf{u}) \right)^2 \de_{\ell}\mathbf{u}.
$$
In this context, the expected value $\lambda(\mathbf{u})$ is derived from the direct computed estimate, i.e., considering each point's bandwidth. The estimated value $\hat{\lambda}(\mathbf{u})$, on the other hand, reflects the approximation derived from partitioned bandwidths, culminating in a rigorous assessment of the intensity estimation's fidelity and reliability.

\begin{figure}[h!tb]
\centering
\includegraphics[width=0.99\linewidth]{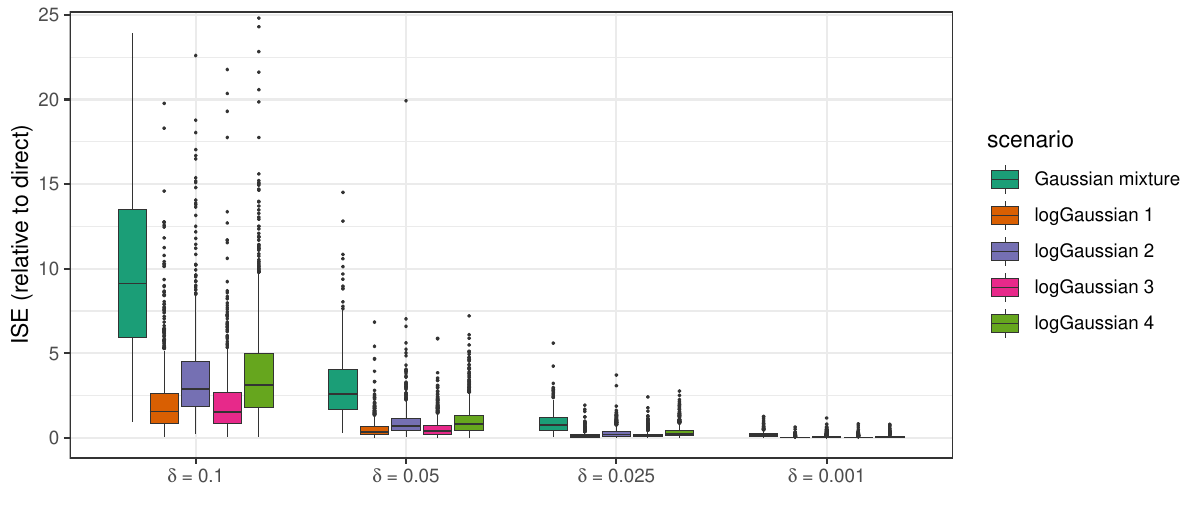}
\caption{Integrated square errors with respect to the direct estimation of intensity function. We consider several values for the parameter $\delta$. 
The resolution of the intensity arrays is $128\times 128$. We truncate the most extreme outliers for better visualisation.}
\label{fig:ISE}
\end{figure}
In Figure \ref{fig:ISE}, the boxplots illustrate the Integrated Squared Errors (ISE) corresponding to various estimates across different bandwidth partitions. This observation aligns with findings from \cite{Davies2018kernel}, where coarser bandwidth partitions tend to yield higher errors. Notably, even with the coarsest partition configuration $(\delta = 0.1)$ involving ten spatial bins, the resulting estimate remains remarkably accurate regarding ISE. As anticipated, as the bandwidth partitions become finer, the estimates converge towards the direct estimation, demonstrating the anticipated trend of improved accuracy with an increased number of bandwidth bins.

\begin{figure}[h!tb]
\centering
\includegraphics[width=0.99\linewidth]{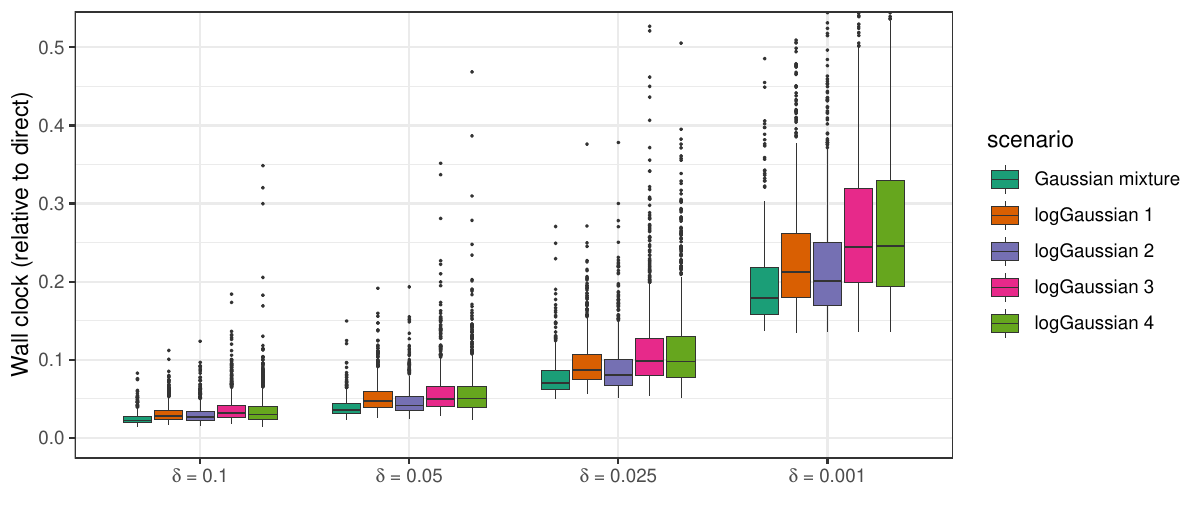}
\caption{Wall clock timings presented in relation to the direct estimation approach, presenting the efficiency gains achieved through the partitioning algorithm.}
\label{fig:times}
\end{figure}

We consider total elapsed execution times relative to direct estimation, i.e., $t_{\text{partition}}/t_{\text{direct}}$. In Figure \ref{fig:times}, a clear correlation emerges between elapsed execution times and the number of bandwidth bins employed. 
Notably, the observed increase in execution times parallels the number of bins. These times consistently remain significantly lower than the time required for direct estimation. This means that the binning procedure offers a much more efficient solution regarding execution times. It is worth noting that when the bandwidth bin count reaches its maximum $(\delta = 0.001)$, there is a notable elongation in processing time. The incremental gains in error reduction do not appear to justify the increase in execution time in this scenario.

\section{Application: Traffic accidents in some Colombian cities}\label{sec:application}
Medellin-Colombia is known for its unique division by {\it comunas}. These comunas are administrative divisions that classify and organise the city into distinct neighbourhoods. Medellin is divided into 16 comunas with distinctive characters, history, and social dynamics. The bustling urban hub of {\it Comuna 10} is popularly known as {\it La Candelaria}. The division by comunas allows for focused governance and tailored approaches to address the specific needs of each neighbourhood, fostering community engagement and empowerment. 

A compilation of all traffic accidents transpiring between 2014 and 2019 in Medell\'in was undertaken, with Candelaria being selected as the focal subnetwork. This choice was deliberate, as Candelaria lies within the city's central region, where most accidents occur, and its structural simplicity renders it a fitting candidate for analysis. The dataset containing records of traffic accidents within Candelaria is publicly available on the OpenData portal maintained by the Medellin Town Hall. Additionally, the shapefile depicting the intricate road network of the area is also accessible for reference and analysis \citep{SMM}. In Medellin's Candelaria comuna, traffic accidents remain a pressing issue, as evidenced by recent reports and statistics. According to local news sources such as {\it El Colombiano}, this area has notably increased traffic accidents. The combination of high population density and heavy traffic flow contributes to the heightened risk of collisions. The narrow and congested streets and the presence of informal vendors and pedestrians further compound the problem. Notably, motorcycles, which are popular means of transportation in the city, have been involved in a significant number of accidents. The city authorities have taken steps to address the issue, including implementing stricter traffic regulations and improving infrastructure. However, sustained efforts are required to mitigate the frequency and severity of traffic accidents in Medellin's Candelaria zone.

\begin{figure}[h!tb]
	\centering
	\begin{subfigure}{.48\linewidth}
		\centering
		\includegraphics[width=0.99\linewidth]{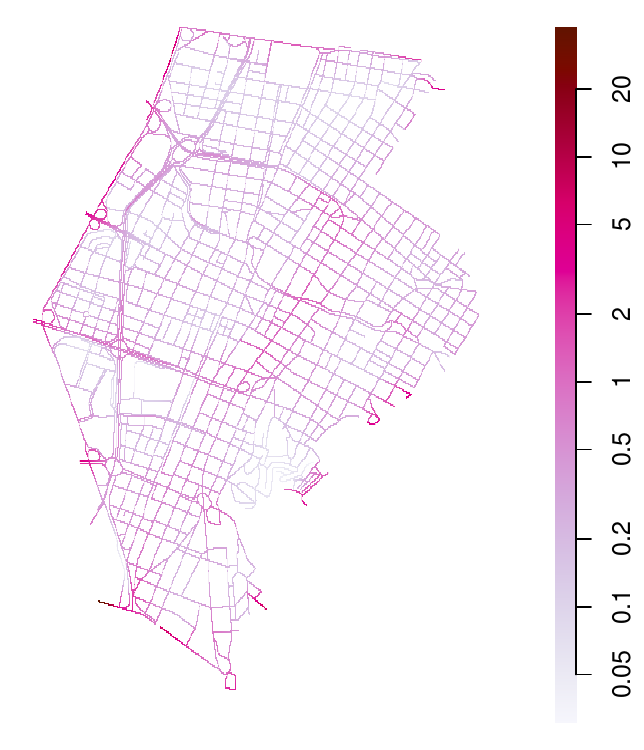}
		\caption{}
	\end{subfigure}
	\begin{subfigure}{.48\linewidth}
		\centering
		\includegraphics[width=0.99\linewidth]{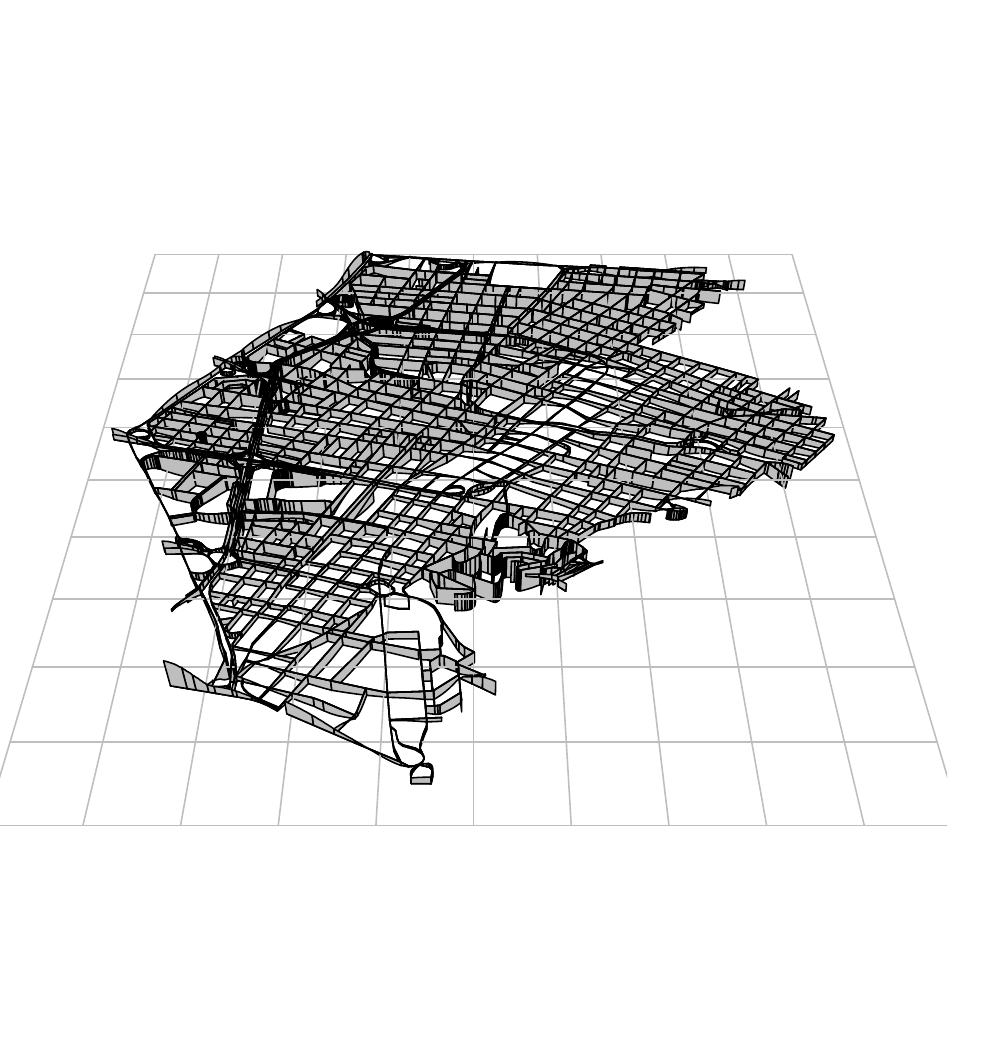}
		\caption{}
	\end{subfigure}
	\caption{Estimated adaptive heating kernel intensity using the partition algorithm for Candelaria's traffic accidents. (a) The pixel intensity values are displayed as colours as a pixel image on the linear network. (b) Estimated intensity given as a perspective view.
}
	\label{fig:adaptivecandelaria1}
\end{figure}

To ascertain the expected number of traffic accidents within Candelaria through the heating kernel in its adaptive version, applying the partitioning algorithm introduced in Section \ref{sec:partitioningalgorithm} was necessary. A total of 276 distinct groups $(\approx \sqrt{76610})$ were selected to determine the bandwidth bins for the analysis, where $76610$ is the total number of points in the point pattern. Figure \ref{fig:adaptivecandelaria1} displays the estimation result.
The figure illustrates the spatial distribution of traffic accidents in the Candelaria neighbourhood. It becomes apparent that the highest concentration of traffic incidents predominantly occurs in the western region. Notably, the roadways hosting more roundabouts, indicative of larger and busier thoroughfares, exhibit a higher incidence of accidents. Intriguingly, an additional concentration of accidents emerges in the central-eastern portion of Candelaria, hinting at localised factors or traffic dynamics that warrant further investigation. This visualization provides valuable insights into the geographic patterning of accidents within the neighbourhood, offering a foundation for targeted interventions and traffic safety improvements. The execution of the estimation procedure, a computationally complex task, took seven minutes when executed on a conventional computing platform.

\section{Discussion}\label{sec:discussion}
This paper uses variable-width kernels to address the non-parametric intensity estimation challenge of a point process within a linear network. The distinctive feature of these kernels lies in their adaptability—each point in the point pattern is endowed with a bandwidth that varies according to the density of surrounding points. This ensures a finer bandwidth for densely populated areas and a broader one for isolated points, effectively capturing the local heterogeneity of the process. While this technique has been discussed in the literature at a general level, our contribution lies in extending a partitioning algorithm, enabling efficient estimation with minimal information loss for the heating kernel in the linear network context.

We have employed the heat kernel method, a sophisticated approach. Nevertheless, other techniques, such as Equal-split kernel estimators, have proven reliable in the fixed bandwidth case. We explored the adaptive version of Equal-split estimators but found them computationally infeasible. On the other hand, adaptive versions of Edge-corrected estimators are plausible and conceivable, though some researchers have already ventured into this domain.

In our simulations, we have chosen to model fields, whether random or not, within a planar region, namely the unit square, and subsequently constrained these fields to our linear network embedded within the unit square.
This approach serves as a quick and straightforward means of simulation within the linear network. However, it is important to note that contemporary methods now enable simulations directly on the network of random fields \citep[see, e.g., \ ][]{bolin2023gaussian}.

Our adaptive kernel-based technique can be extended to risk estimation. In our practical example, this could involve estimating the relative risk of accidents in Candelaria based on the known vehicle intensity. Identifying areas with elevated accident risk, even if not the most intense, could facilitate collaboration with authorities to establish surveillance systems.

Another avenue for expansion is the incorporation of time. In cases where each point's occurrence is linked to a temporal dimension, an adaptive estimation with an additional dimension becomes necessary. Certainly, there are various avenues for future exploration, such as addressing networks with curvature, which poses a challenge for adaptive estimation and traditional estimation methods.

Finally, our research has yielded a swift and effective technique for estimating the intensity of point processes within linear networks, particularly in scenarios marked by high heterogeneity. This paper contributes to the analysis of point patterns in linear networks by providing practical insights into adaptive estimation methods, underlining their significance in addressing real-world spatial phenomena.

\bibliographystyle{dcu}
\bibliography{bibliography}

\end{document}